\documentclass[twocolumn]{aastex631}

\usepackage{graphics,epsf}
\usepackage[utf8]{inputenc}
\usepackage{amsmath}                % American Mathematical Society package
\usepackage{amsfonts}               % American Mathematical Society fonts
\usepackage{amssymb}                % American Mathematical Society symbol
\usepackage{epsfig}                 % EPS figures
\usepackage{graphicx}
\usepackage{float}
\usepackage{color}
\usepackage{multirow}               % double row table entries
\usepackage{hyperref}

\hypersetup{
    colorlinks=true,
    linkcolor=red,   
    urlcolor=cyan}

\usepackage[para,online,flushleft]{threeparttable}

\usepackage[colorinlistoftodos]{todonotes}

\newcommand{\mesa}{\textsc{mesa}}

\newcommand{\kms}{{~\rm km\; s^{-1}}}

\newcommand{\km}{{~\rm km}}
\newcommand{\s}{{~\rm s}}

\newcommand{\g}{{~\rm g}}
\newcommand{\G}{{~\rm G}}
\newcommand{\K}{{~\rm K}}
\newcommand{\erg}{{~\rm erg}}

\begin{document}

\title{The implications of large binding energies of massive stripped core collapse supernova progenitors on the explosion mechanism}
\date{January 2023}

\author[0000-0002-9444-9460]{Dmitry Shishkin}
\affiliation{Department of Physics, Technion, Haifa, 3200003, Israel; s.dmitry@campus.technion.ac.il; soker@physics.technion.ac.il}

\author[0000-0003-0375-8987]{Noam Soker}
\affiliation{Department of Physics, Technion, Haifa, 3200003, Israel; s.dmitry@campus.technion.ac.il; soker@physics.technion.ac.il}

\begin{abstract}
We examine the binding energies of massive stripped-envelope core collapse supernova (SECCSN) progenitors with the stellar evolution code \mesa, and find that the jittering jets explosion mechanism is preferred for explosions where carbon-oxygen cores with masses of $\ga 20 M_\odot$ collapse to leave a neutron star (NS) remnant. We calculate the binding energy at core collapse under the assumption that the remnant is a NS. Namely, stellar gas above mass coordinate of $\simeq 1.5-2.5 M_\odot$ is ejected in the explosion. We find that the typical binding energy of the ejecta of stripped-envelope progenitors with carbon-oxygen core masses of $M_{\rm CO} \ga 20 M_\odot$ is $E_{\rm bind} \ga 2 \times 10^{51} \erg$. We claim that jets are most likely to explode such cores as jet-driven explosion mechanisms can supply high energies to the explosion. We apply our results to SN 2020qlb, which is a SECCSN with a claimed core mass of $\simeq 30-50 M_\odot$, and conclude that the jittering jets explosion mechanism best accounts for such an explosion that leaves a NS. 
\end{abstract} 


% ==========================================================
\section{INTRODUCTION}
\label{sec:Intro}
% ==========================================================
% cite: author (year) | citep: (author year) | citealt: author year

The binding energy of a core collapse supernova (CCSN) progenitor plays a crucial role in determining the explosion outcome, like explosion energy and remnant mass. Typical explosion energies are estimated to be in the range of $E_{\rm exp} \simeq 10^{50} - 10^{52} \erg$ (e.g., \citealt{YangChevalier2015_crabEnergy, Utrobin_etal_2015_1987Aenergy, GalYam2019_reviewLuminous, Burrows2022review}).
The binding energy of the most massive pre-collapse cores have similar or even larger values than these typical explosion energies (e.g., \citealt{PejchaThompson_2015_progenitorsBinding, Bruenn_etal_2016, Chan_etal_2020, WangVartanyanBurrowsColeman2022, Burrows_etal_2020_simulations}). The explosion mechanism should both overcome the binding energy and account for the explosion energy (radiation + final kinetic energy of the ejecta). 

Two theoretical explosion mechanisms of non-rotating (or slowly rotating) pre-collapse cores utilize the gravitational energy of the collapsing core to power CCSNe. These are the delayed-neutrino explosion mechanism (e.g., \citealt{BetheWilson1985, Ertl_etal_2016, Burrows_etal_2020_simulations, Bruenn_etal_2020_neutrinoExplosionCode, Bollig_etal_2021, BurrowsVartanyan_2021_Nature, Zhaetal2023}) and the jittering-jets explosion mechanism (e.g., \citealt{Soker2010, Papish2011_jittering, GilkisSoker2015, Soker2019_SASIjets, Soker2022_Boosting}). 
Studies show that the maximum energy that the delayed neutrino mechanism can supply to overcome the binding energy of the ejecta is $E_{\nu}^{\rm max} \simeq 2\times 10^{51} \erg$, resulting in maximum explosion energies (after removing the binding energy) of $E_{\rm exp} \simeq 2 \times 10^{51} \erg$ (e.g., \citealt{Fryer_etal_2012, Ertl_etal_2016, Sukhbold_etal_2016, Gogilashvili_etal_2021}). 
In addition, the delayed neutrino mechanism has problems in producing the observed amount of $^{56}{\rm Ni}$, in particular in stripped-envelope CCSNe (SECCSNe; for a recent study see \citealt{SawadaSuwa2023})

The above limits on the explosion energies of the delayed neutrino mechanism come either from direct calculations (e.g., \citealt{Bollig_etal_2021, Gogilashvili_etal_2021}) or by scaling to observed CCSNe, such as the Crab nebula and SN~1987A (e.g., \citealt{Sukhbold_etal_2016}). We do note that some studies do attribute larger explosion energies to the delayed neutrino mechanism, e.g., \cite{Fryeretal2018} who base their explosion energies on \cite{Fryer_etal_2012}. \cite{Fryer_etal_2012} adopt the convection-enhanced neutrino-driven paradigm and assume that the explosion energy is equal to the energy stored in the convective region at the time of collapse, and derive a maximum available energy of $3-4 \times 10^{51} \erg$. However, they derive these high values of explosion energies with progenitor masses of $\la 20 M_\odot$ only, while we focus on more massive progenitors. We therefore consider here the results of the later studies that we quoted above that give the limit of $E_{\nu}^{\rm max} \simeq 2\times 10^{51} \erg$. In any case, this discussion suggests that we should take this upper limit of the explosion energy in the delayed neutrino mechanism with caution.

The jittering jets explosion mechanism, on the other hand, can account for much larger explosion energies (e.g., \citealt{GilkisSokerPapish2016, Soker2022review}). The magneto rotational explosion mechanism that works only for rapidly rotating pre-collapse cores can also account for large explosion energies (e.g., \citealt{LeBlancWilson1970, Khokhlov_etal_1999, LopezCamara2013, Wheeler_etal_2015, BrombergTchekhovskoy2016, Kuroda_etal_2020_magnetorotational, Gottlieb_etal_2022_transients, Gottlieb_etal_2022_collapsar, Fujibayashi_etal_2022, Jade_etal_2022}) because the newly born neutron star (NS) launches fixed-axis jets.
Whether the explosion is by jittering jets (most CCSNe according to that model; e.g., \citealt{Soker2022review}) or by fixed-axis jets, the jets operate in a negative feedback cycle, i.e., the jet feedback explosion mechanism (e.g., \citealt{Soker2016Review}). As well, the jets can influence the direction of later jets (e.g., \citealt{PapishSoker2014, Gottlieb_etal_2022_wobbly}). 
Even if the stochastically accreted mass has sub-Keplerian angular momentum, it might still form an accretion belt that can launch jets (e.g., \citealt{SchreierSoker2016, GarainKim2022}).

The above discussion implies that CCSNe with large kinetic energies of $E_{\rm exp} \ga 2 \times 10^{51}$, e.g., SN~2020jfo (\citealt{SN2020jfo} -- $E_{\rm exp} = 2.9 \times 10^{51} \erg$, but see \citealt{SN2020jfo_Tejaetal2022} for a lower estimated energy); SN~2020qlb (\citealt{SN2020qlb_magneto_32_48} -- $E_{\rm exp} = 20 \times 10^{51} \erg$); SN~2012au (\citealt{SN2012au} -- $E_{\rm exp} = 4.8-5.4 \times 10^{51} \erg$) require jets to drive the explosion. 

In this study we examine the interesting case of the hydrogen-poor and super-energetic CCSN SN~2020qlb. \cite{SN2020qlb_magneto_32_48} estimate the kinetic energy of SN~2020qlb as $\simeq 20 \times 10^{51} \erg$ and suggest that a magnetar supplies the large amount of energy of the ejecta. They also provide fitting parameters for the magnetar model (e.g., \citealt{Maeda_etal_2007, KasenBildsten2010, Woosley2010, Metzger_etal_2015, Nicholl_etal_2017,Gomez_etal_2022}) and estimate the ejecta plus the remnant mass to be $M_{\rm ej}+M_{\rm rem} \simeq {\rm few} \times 10 M_\odot$. We consider a stripped envelope supernova, i.e., the envelope mass is very low. Therefore, the pre-collapse core mass is the sum of the ejecta mass, the remnant mass, and a small amount of mass carried as energy by neutrinos: $M_{\rm core} \simeq M_{\rm ej}+M_{\rm rem} \simeq  {\rm few} \times 10 M_\odot$.

It seems nonetheless, that there are two reasons why the explosion of SN~2020qlb must be driven by jets and not by the delayed neutrino mechanism. 
The first one is that the formation of an energetic magnetar must be accompanied by the launching of even more energetic jets at the explosion itself and possibly after the explosion as well (e.g. \citealt{Soker2016magnetar, Soker2017magnetar, SokerGilkis2017magnetar, Shankaretal2021magnetar, Soker2022LSNe, Soker2022review}). 
The second reason is the new finding of the present study, where by a stellar evolutionary code (section \ref{sec:NumScheme}) we show (section \ref{sec:Results}) that the binding energies of such massive cores are above the energies the delayed neutrino mechanism can supply \citep{Janka2012review,Soker2022review}.
In section~\ref{sec:DiscussionSummary} we summarize our results and discuss their implications in the context of the jet feedback explosion mechanism.  

% ==========================================================
\section{Numerical Scheme}
\label{sec:NumScheme}
% ==========================================================

We use the stellar evolution code \mesa~(\citealt{Paxtonetal2010,Paxtonetal2013,Paxtonetal2015,Paxtonetal2018,Paxtonetal2019}) to simulate the structure of $52$ CCSN progenitor models, all with initial metalicity of $z=0.02$. 
We base our numerical routine on the `20M\_pre\_ms\_to\_core\_collapse' example from \mesa~r22.05.1, but simulate the evolution using version r15140. Starting from the zero age main sequence (ZAMS), we evolve the star until center He depletion (when helium abundance in the center is either $\simeq 1\%$ or $\simeq 5\%$), at which point we numerically remove the hydrogen-rich envelope, leaving only the He core. We then evolve the star until core-collapse.
We introduce several changes to the example routine to both adhere to our requirements, i.e., having at least 21 isotopes and having sufficiently high resolution \citep{ShishkinSoker2021}, and to fit \mesa~version r15140. We expand on the numerical scheme in appendix~\ref{App1:num_pres}.

By varying the ZAMS stellar mass, the wind parameters, and the exact time when we numerically remove the envelope we obtain stripped-envelope CCSN progenitors, i.e., hydrogen-poor stellar progenitors, with varying mass of a carbon-oxygen (CO) core, $M_{\rm CO} \simeq 4-35 M_\odot$. This mass range of SECCSNe corresponds to ZAMS stellar masses range of $\simeq 20-82 M_\odot$. In some cases the core contains several solar masses of helium, while in other cases the core has a much lower helium mass, depending on the above parameters. 
 
Because the inner part of the core collapses to form a NS or a black hole (BH) remnant, only the binding energy of the outer core is relevant to our study. We calculate the binding energy of the outer core, $E_{\rm bind}(r)$, by integrating over the sum of the internal energy and gravitational energy from the surface down to the mass coordinate that separates the ejecta and the final remnant $m=M_{\rm in}$, i.e., the inner boundary of the ejecta. We calculate the binding energy of the ejecta for two values of this mass coordinate $M_{\rm in}=1.5~,2.5 M_\odot$, because we consider cases where the remnant is a NS.  

We refer to the latest evolution point we simulate in the stellar evolution as collapse. This point of collapse must adhere to an iron core more massive than $M_{\rm Fe}>1.4$, where we assume collapse is imminent. This value of iron core mass is more or less when the iron core mass reaches its maximum value (as at the onset of collapse the iron disintegrates). About half of simulations reach infall velocities at the edge of the iron core of $v_{\rm fe,infall}>100 \kms$. Other simulations encounter numerical difficulties and we had to terminate them at somewhat earlier times. 

% ==========================================================
\section{Results}
\label{sec:Results}
% ==========================================================
% ==================================
\subsection{Binding Energy towards Collapse}
\label{subsec:Scaling}
% ==================================

\begin{figure}[t]
\begin{center}
\includegraphics[trim=0.5cm 0cm 0.5cm 0.57cm,scale=.6]{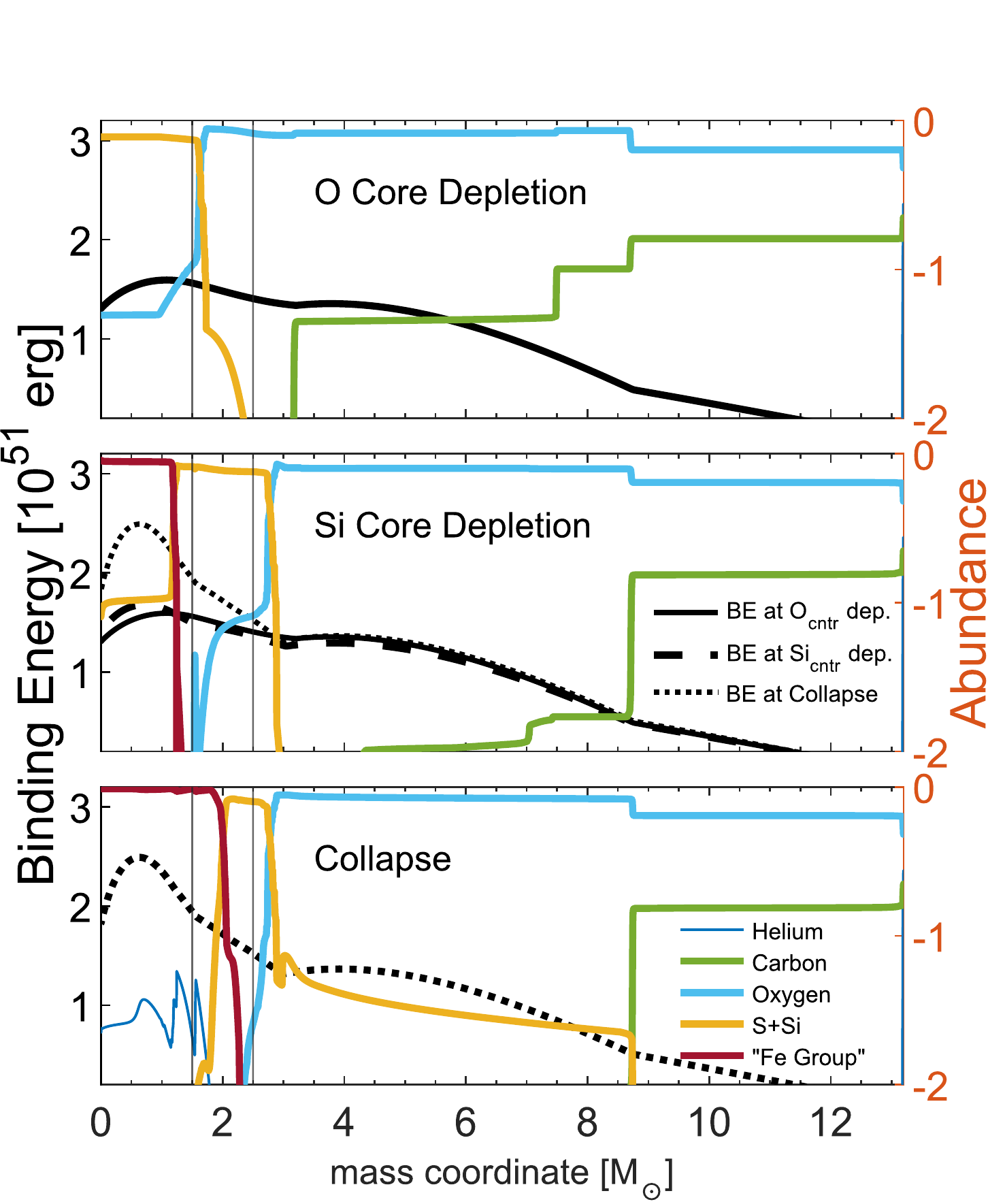} %[trim=left lower right upper]
\caption{Abundances and the binding energy as function of mass coordinate at three times for a SECCSN (hydrogen-poor) progenitor model with carbon-oxygen core mass of $M_{\rm CO}^{\rm collapse} = 13.2 M_\odot$, which corresponds to a ZAMS mass of $M_{\rm ZAMS} \approx 40 M_\odot$. 
The colored step-like lines are the abundances according to the inset in the lower panel. The black and smoothly varying lines represent $E_{\rm bind}(m)$, which is the binding energy of the envelope laying above mass coordinate $m$.
The three panels present these quantities at three different times: center oxygen depletion (upper panel, binding energy by the black solid-line), center silicon depletion (middle panel; binding energy by the black dashed-line), core collapse (lower panel; binding energy by black dotted-line).  Note that the middle panel contains the binding energy at the three times to allow for comparison. The two vertical lines mark the mass coordinates $m=M_{\rm in}=1.5 M_\odot$ and $m=M_{\rm in}=2.5 M_\odot$.
Helium that appears only at collapse results from disintegration of iron.}
\label{fig:BindingEnergyEvolve1}
\end{center}
\end{figure}

As the stellar core evolves towards collapse and nucleosynthesis of heavier elements takes place, the core becomes denser and the binding energy of the inner layers of the star increases. We demonstrate this for one stellar model of carbon-oxygen core mass (mainly oxygen) of $M_{\rm CO}=13.2 M_\odot$ in Fig~\ref{fig:BindingEnergyEvolve1} where we present the star at three times: at center oxygen depletion, at center silicon depletion, and at collapse. We present the composition of the main isotopes by lines with different colors and the binding energy $E_{\rm bind}(m)$ by the black lines. Here $E_{\rm bind}(m)$ is the binding energy (gravitational + internal) of the envelope laying above mass coordinate $m$. 
Relevant to this study is the binding energy of the ejecta, which is the mass above mass coordinate $m=M_{\rm in} \simeq 1.5,~2.5 M_\odot$. The mass $M_{\rm in}$ is the baryonic mass of the NS remnant (the corresponding final gravitational masses will be $\simeq 1.35,~2.1 M_\odot$). We mark these two masses by the vertical lines. We see that at collapse the binding energy $E_{\rm bind} (M_{\rm in})$ is larger than at earlier times. 
% FFFFFFFFFFFFFFFFFFFFFFFFFFFFFFFFFFFFFFFFFFFFFFFFFFFFFFFFFFFFF
% FFFFFFFFFFFFFFFFFFFFFFFFFFFFFFFFFFFFFFFFFFFFFFFFFFFFFFFFFFFFF

\begin{figure} [t!]
\begin{center}
\includegraphics[trim=0.5cm 0cm 0.5cm .57cm,scale=.6]{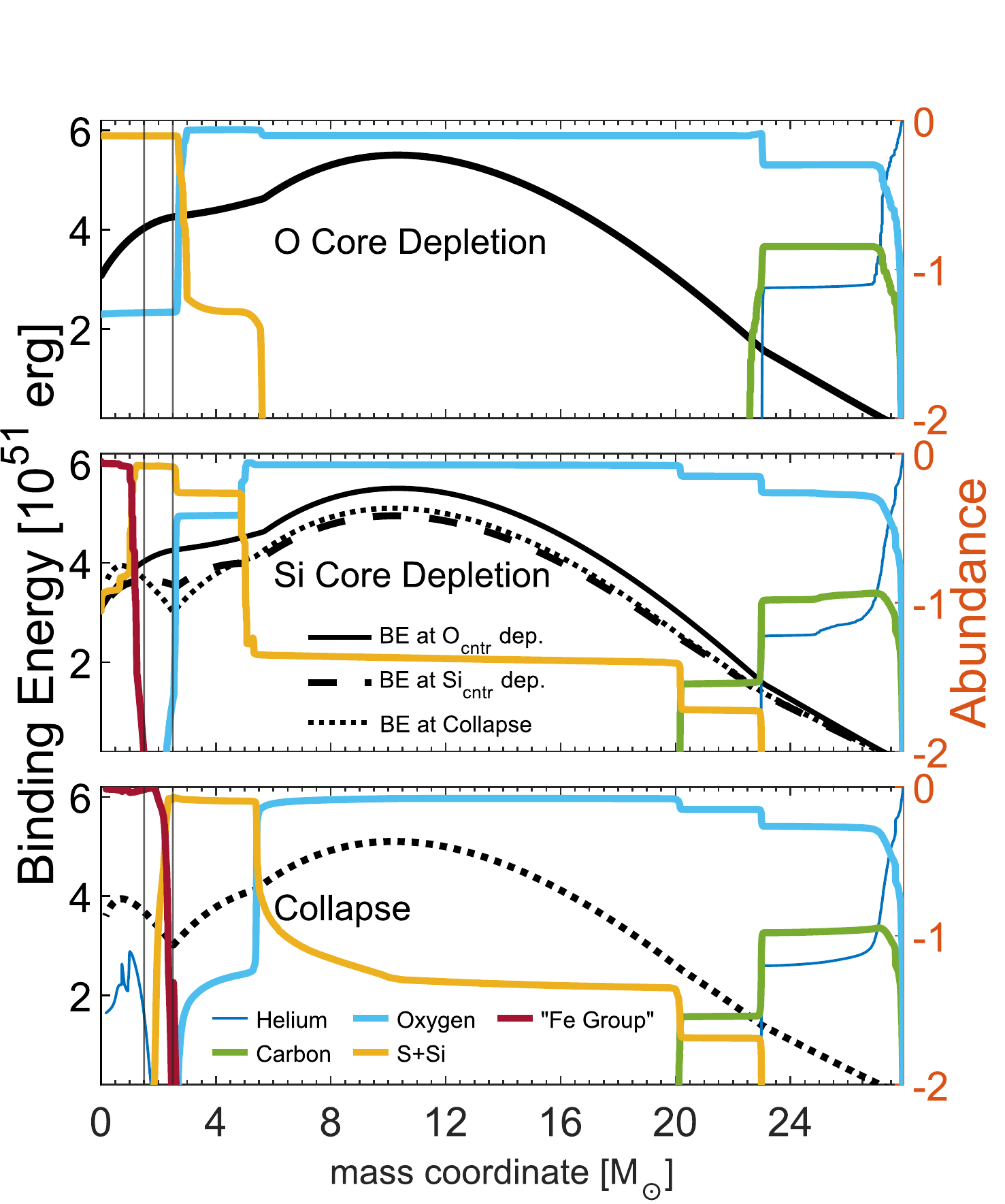} %[trim=left lower right upper] %8.9cm 8cm 8.5cm 8cm for old save
\caption{Similar to Fig.~\ref{fig:BindingEnergyEvolve1} but for a more massive core of $M_{\rm CO}^{\rm collapse} = 26.5 M_\odot$, which corresponds to a ZAMS mass of $M_{\rm ZAMS} \approx 65 M_\odot$. Note that the left vertical axis is scaled differently than in Fig.~\ref{fig:BindingEnergyEvolve1}.}
\label{fig:BindingEnergyEvolve2}
\end{center}
\end{figure}

In Fig. \ref{fig:BindingEnergyEvolve2} we present composition and binding energy for a model with a much more massive core of $M_{\rm CO}=27 M_\odot$. There are two qualitative differences between this model and the one we present in Fig. \ref{fig:BindingEnergyEvolve1}. The first qualitative difference is that the binding energy at collapse is somewhat smaller than at the earlier time that we present in the figure. The explanation to the decreasing binding energy shortly before collapse is that the envelope expands starting from deep in the oxygen burning shell and outwards. We find (by drawing the density profiles) that moving from the upper to the middle panel of Fig. \ref{fig:BindingEnergyEvolve2} the density from $m \simeq 10 M_\odot$ and outward decreases, reducing the binding energy. This mass coordinate is deep inside the shell where oxygen (teal line) burns to S+Si (yellow line).  
% FFFFFFFFFFFFFFFFFFFFFFFFFFFFFFFFFFFFFFFFFFFFFFFFFFFFFFFFFFFFF
% FFFFFFFFFFFFFFFFFFFFFFFFFFFFFFFFFFFFFFFFFFFFFFFFFFFFFFFFFFFFF

The second qualitative difference comes from the much higher binding energy of the ejecta of the descendant CCSN of the more massive model, i.e., $E_{\rm bind} (M_{\rm in}) \ga 2 \times 10^{51} \erg$. The implication is that we do not expect that the neutrino driven explosion mechanism can account for explosions of such cores. We argue that jets explode these cores. We leave the discussion of this point, as well as our view that jets also explode cores with lower binding energy, to section \ref{sec:DiscussionSummary}, where we also refer to the claim of a very massive core of SN~2020qlb \citep{SN2020qlb_magneto_32_48}. We first find the range of such high-binding-energy cores. 

% ==================================
\subsection{High-binding-energy cores}
\label{subsec:Binding}
% ==================================

\begin{figure*}
\begin{center}
\includegraphics[trim=1.5cm 0cm 0cm .57cm,scale=0.65]{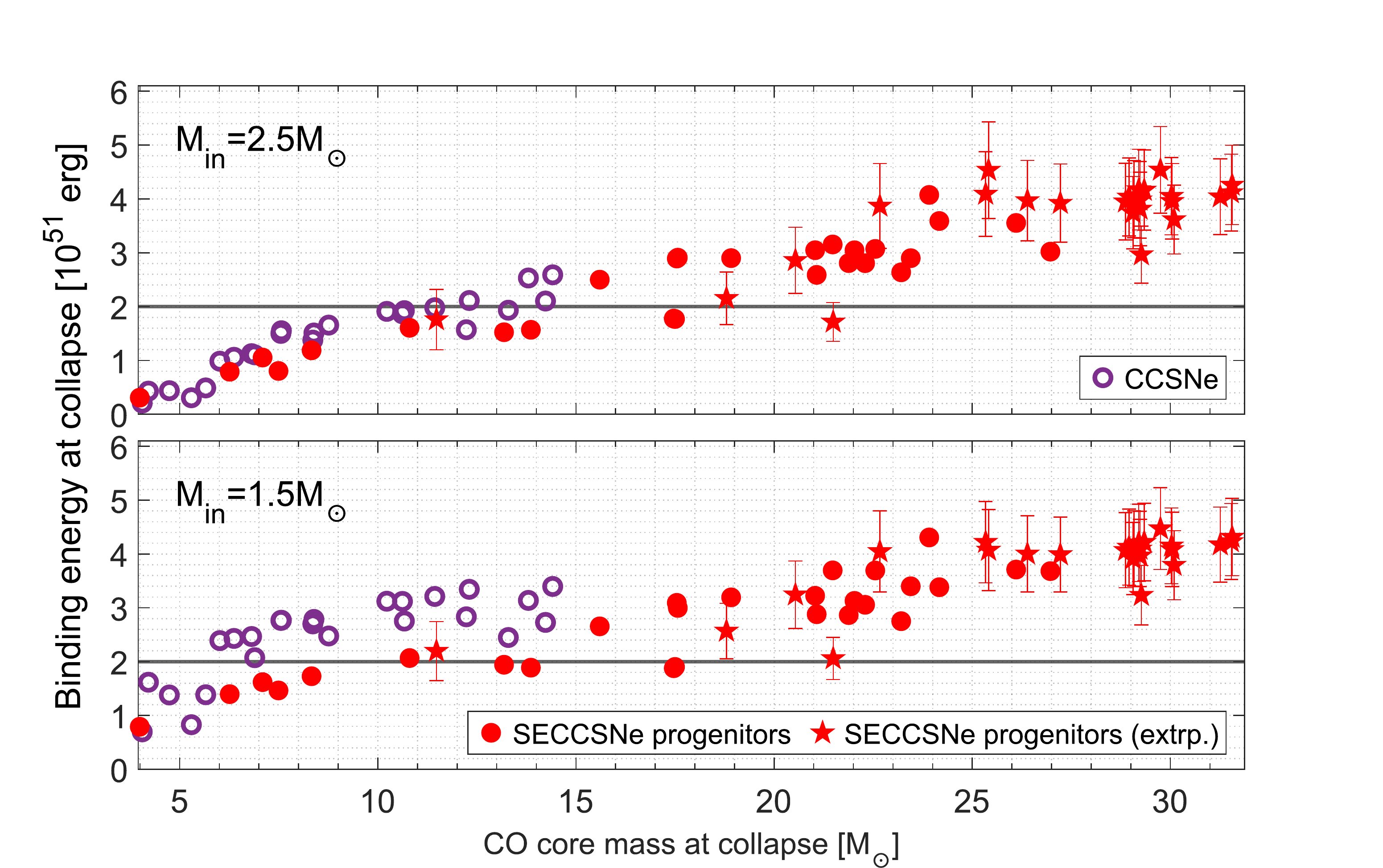} %[trim=left lower right upper] %1.2cm 4.2cm 1.8cm 4.38cm
\caption{The binding energies of the simulated models as a function of the carbon-oxygen core mass nearing collapse (vertical axis). The panels show the final binding energy at two mass coordinates: $M_{\rm in}=2.5M_\odot$ (top) and $M_{\rm in}=1.5M_\odot$ (bottom). The red data points (filled circles and stars at the outer panels) are stripped-envelope (SE) models (SECCSNe), whilst purple empty-circles data points are models from \cite{ShishkinSoker2022} that have hydrogen rich envelopes. Red stars are extrapolated data points, as explained in appendix \ref{App2:BEcalc}. The horizontal line at $2 \times 10^{51} \erg$ denotes the binding energy above which we do not expect the neutrino delayed explosion mechanism to explode the core.}
\label{fig:BindingEnergy}
\end{center}
\end{figure*}

We search for the mass range of cores that have binding energies at collapse of $E_{\rm bind} (M_{\rm in}) \ga 2 \times 10^{51} \erg$. 
We present the results in Fig. \ref{fig:BindingEnergy}.
We present the binding energy for an inner ejecta mass coordinate of $M_{\rm in} =2.5 M_\odot$ (upper panel) and $M_{\rm in} =1.5 M_\odot$ (lower panel).
We focus on the binding energy of these two mass coordinates $M_{\rm in} = 1.5 - 2.5 M_\odot$ as the iron core masses at collapse falls within this mass range.
The horizontal line at $2 \times 10^{51} \erg$ is the approximate energy above which we do not expect that neutrino heating by itself can explode the core. In appendix~\ref{App2:BEcalc} we provide linear fits to the binding energy at collapse as function of CO core mass for these two mass coordinates $M_{\rm in} = 1.5,~2.5 M_\odot$ (table~\ref{tab:BElines}).
% FFFFFFFFFFFFFFFFFFFFFFFFFFFFFFFFFFFFFFFFFFFFFFFFFFFFFFFFFF
% FFFFFFFFFFFFFFFFFFFFFFFFFFFFFFFFFFFFFFFFFFFFFFFFFFFFFFFFFF

We simulated $52$ stripped-hydrogen envelope cases. In $27$ cases the cores reach collapse as we present by the red circles in the figure. In $25$ cases cases the numerical code encountered difficulties and we had to stop the simulation before reaching collapse. In these cases we extrapolate from an evolutionary time before collapse to the collapse time as we explain in appendix~\ref{App2:BEcalc} (red stars in the figure). We also include in the figure the binding energies of models with hydrogen-rich envelope that we take from \cite{ShishkinSoker2022}, as we mark by open purple circles. The values we calculate for the binding energies of these core-collapse progenitors are similar to those of other studies (e.g., \citealt{ WangVartanyanBurrowsColeman2022, Burrows_etal_2020_simulations}).

From Fig. \ref{fig:BindingEnergy} (with a more rigorous derivation in appendix~\ref{App2:BEcalc}) we draw our conclusion that in cases where the inner mass of $M_{\rm in} =2.5 M_\odot$ of the core collapses to form a NS, the delayed neutrino mechanism cannot explode cores with masses of $M_{\rm CO} \ga 15 M_\odot$ (or maybe rarely do so). For $M_{\rm in} =1.5 M_\odot$ we find this limit to be $M_{\rm CO} \ga 13 M_\odot$.

\begin{figure*}
\begin{center}
\includegraphics[trim=1cm 0cm 0cm 0.65cm,scale=0.63]{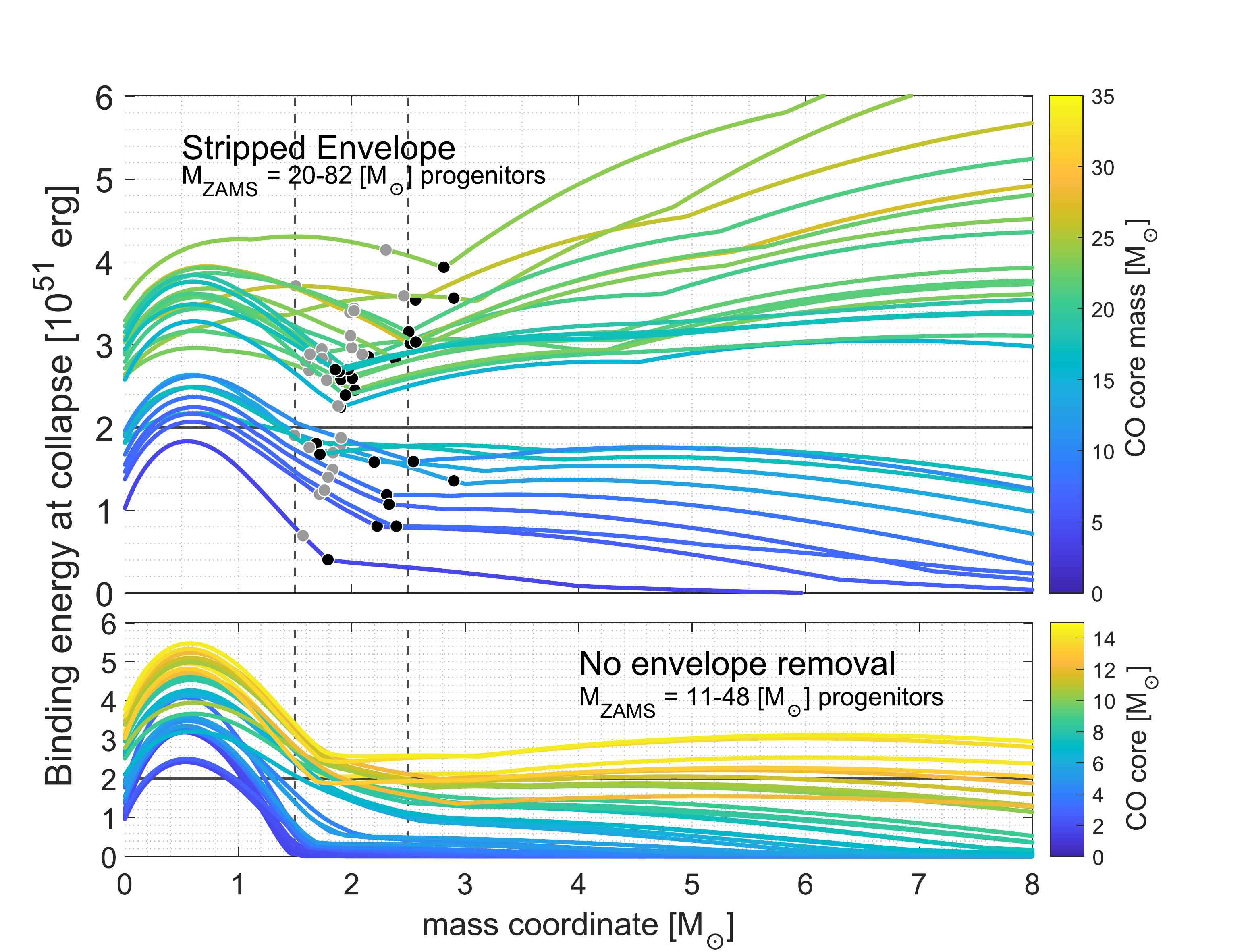} %[trim=left lower right upper] %1.2cm 4.2cm 1.8cm 4.38cm
\caption{Upper panel: The binding energies of the simulated stripped-envelope models (vertical axis) as a function of the mass coordinate (horizontal axis) at collapse. The color of each line signifies the final CO core mass at collapse, scaled with the color bar on the right. Black dots mark the mass coordinate where there is a steep decrease in the density inside the core. Gray squares denote the edge of the iron core. Lower panel: The binding energy for hydrogen-rich envelope models from \cite{ShishkinSoker2022}. The grey horizontal line marks more or less the upper binding energy that the neutrino delayed mechanism can account for. }
\label{fig:BindingEnergyDetailed}
\end{center}
\end{figure*}

In Fig. \ref{fig:BindingEnergyDetailed} we present a more detailed binding energy profile of the pre-collapse stripped-envelope models that we simulated. When we take the lowest binding energy of the inner core at the onset of collapse we find the lower limit of core masses that the neutrino-driven explosion cannot account for to be $M_{\rm CO} \simeq 15 M_\odot$. 
However, due to possible uncertainties in the calculations by \textsc{mesa}, evident by the spread of points in Fig. \ref{fig:BindingEnergy} and the different binding energies of similar core mass models in Fig. \ref{fig:BindingEnergyDetailed}, we conservatively take the limit at $M_{\rm CO} \simeq 20 M_\odot$. Namely, we argue that jets are responsible for the explosions of all cores of masses $M_{\rm CO} \ga 20 M_\odot$.
% FFFFFFFFFFFFFFFFFFFFFFFFFFFFFFFFFFFFFFFFFFFFFFFFFFFFFFFFFF
% FFFFFFFFFFFFFFFFFFFFFFFFFFFFFFFFFFFFFFFFFFFFFFFFFFFFFFFFFF

In table~\ref{tab:BElines} we provide the linear fit parameters for the binding energy at collapse for the edge of the iron core (gray squares in the figure) and the binding energy curve break (black circles). The binding energy curve break is point we refer to as separating the inner core from the outer core, and is the point of lowest binding energy for most simulated cases that reached collapse.

% ==========================================================
\section{Discussion and Summary}
\label{sec:DiscussionSummary}
% ==========================================================

We simulated the evolution of $52$ massive SECCSN progenitor models corresponding to ZAMS masses of $20 \la M_{\rm ZAMS} \la 82$. We removed the entire hydrogen-rich envelope, and calculated the binding energy just before core collapse. 
The final core mass depends on the ZAMS mass and on the mass loss parameter (appendix \ref{App3:simList}). We present the structure of the pre-collapse progenitor for two cases in Figs. \ref{fig:BindingEnergyEvolve1} and \ref{fig:BindingEnergyEvolve2}. 
We find that to a fare accuracy we can linearly fit the binding energy of these stripped-envelope progenitors to the CO core mass $M_{\rm CO}$ (Fig.~\ref{fig:BindingEnergy} and table~\ref{tab:BElines}). 

We present our main results in Fig. \ref{fig:BindingEnergy}. In those figures the horizontal gray line represents the approximate maximum energy that the neutrino-driven mechanism can supply, $E_{\nu}^{\rm max}=2\times 10^{51} \erg$. We find that the binding energy calculated at $M_{\rm in}=1.5 M_{\odot}$ and $M_{\rm in}=2.5 M_{\odot}$, of progenitors with a carbon-oxygen core mass of $M_{\rm CO} \ga 13 M_\odot$ and $M_{\rm CO} \ga 15 M_\odot$, respectively, are larger than $E_{\nu}^{\rm max}$. 
 Namely, 
 \begin{equation}
E_{\nu}^{\rm max} \la 
    \begin{cases}
        & E_{\rm bind,1.5}  \quad {\rm for} \quad M_{\rm CO} \ga 13 M_\odot 
        \\
        & E_{\rm bind,2.5}  \quad {\rm for} \quad M_{\rm CO} \ga 15 M_\odot .
   \end{cases}
\label{eq:deltaAngle}
\end{equation}

We account for the uncertainty of our simulations and parameter choices by setting a higher upper limit of $\simeq 20 M_\odot$ for the cores that the neutrino explosion mechanism might explode. The main conclusion is that the delayed neutrino explosion mechanism is unlikely to explode stars with a core mass of $M_{\rm CO} \ga 20 M_\odot$. The jittering jets explosion mechanism, on the other hand, has no limiting explosion energy in these ranges as it is fueled by accretion onto the compact remnant (e.g., \citep{GilkisSokerPapish2016, SokerGilkis2017magnetar}).

Let us apply our results to a specific SECCSN. In a recent paper \cite{SN2020qlb_magneto_32_48} 
deduce that SN 2020qlb had an explosion energy of $\simeq 20 \times 10^{51} \erg$ and estimate the progenitor pre-explosion mass, ejecta plus remnant mass, to be $M_{\rm ej}+M_{\rm rem} \simeq  30 -50 M_\odot$. According to our results the binding energy alone of such cores is $E_{\rm bind} > 3 \times 10^{51} \erg$. We therefore conclude that jets must have exploded SN 2020qlb. Jets can also supply the kinetic energy of the ejecta. Namely, jet-driven explosions might make the magnetar powering less critical or not needed at all (although a magnetar might be present). 
If indeed a magnetar, i.e., a rapidly rotating magnetized NS, was formed, then most likely the explosion was via jittering jets. The reason is that an explosion driven by a fixed-axis jets, like if the core is rapidly rotating, will not expel much mass from the equatorial region, which in turn is accreted by the newly formed central object. Therefore, the final mass of the remnant will be large and the remnant will be a BH (see discussion in \citealt{Soker2022review}). 
We note that strong fixed-axis jets might remove some mass even from the equatorial plane, e.g., \cite{Gottlieb_etal_2022_wobbly}. However, they assumed relativistic jets launched by an already formed BH of $4 M_\odot$. The jittering jets explosion mechanism asserts that fixed-axis jets from a NS cannot remove mass from the equatorial plane, at least for several seconds as 2D and 3D simulations show (\citealt{PapishSoker2014_jetFeedback, PapishSoker2014}). The jets might remove mass from the equatorial plane at later times. However, by that time the central star has already grown to a BH by accreting from the equatorial plane. Note that the jets that the papers above simulated are sub-relativistic, initial jet velocity of $10^5 \km \s^{-1}$. This implies that per unit energy the jets have higher momentum than what relativistic jets have.

On a large scope, our study adds to the growing evidence pointing to the major roles that jets play in the explosion, as well as pre-explosion and post-explosion, of CCSNe (for a recent review see \citealt{Soker2022review}).  

\pagebreak

% ===================================================
\section*{Acknowledgments}
% ===================================================
We thank an anonymous referee for helpful comments.
This research was supported by a grant from the Israel Science Foundation (769/20).
\vspace{-0.35\baselineskip}
% ===================================================
\section*{Data availability}
% ===================================================
The data underlying this article are available in Zenodo, at \href{https://doi.org/10.5281/zenodo.7529670}{https://doi.org/10.5281/zenodo.7529670}.

\clearpage

% Appendices
\appendix

\setcounter{figure}{0}                  
\setcounter{table}{0}

% ==========================================================
\section{Numerical prescription details}
\label{App1:num_pres}
% ==========================================================
\renewcommand\thefigure{A.\arabic{figure}}
\renewcommand\thetable{A.\arabic{table}}

Our numerical scheme files (`inlists') are a modified version of the `20\_pre\_ms\_to\_cc' \textsc{mesa} version r22.05.1 `test\_suite' example. We adapted this example to run on \mesa~version r15140 and incorporated certain parameters (e.g., overshooting and mesh resolution) according to our previous works \citep{ShishkinSoker2021,ShishkinSoker2022}.
The full 'inlists' that we used are available online\footnote{\href{https://doi.org/10.5281/zenodo.7529670}{Zenodo:} Modified inlists to reproduce the models. Also included a full simulation list with key parameters and the simulated models at several different time points.}. Here we mention some of the important parameters.

In a similar fashion to our previous papers which focus on the convective profile of the inner layers of massive stars \citep{ShishkinSoker2021,ShishkinSoker2022}, we use the exponential overshooting prescription \citep{Herwig2000} with symmetrical (both `bottom' and `above') and uniform (all burning regions) settings and $f=0.01 , f_0=0.004$ parameters.

We chose the Henyey scheme \citep{Henyey_etal_1965} for mixing length theory (MLT, \citealt{Vitense1958}) with $\alpha_{\rm MLT}=1.5$. 
We also enable the Ledoux criterion \citep{Ledoux1947} and set thermohaline option to `Kippenhahn' with `thermohaline\_coeff = 1' alongside `$\rm alpha\_{semiconvection}=0.01$.

We make use of the 'Dutch' wind loss scheme (e.g., \citealt{Vink2001,NugisLamers2000}), and vary the scaling factor (along with the initial mass) to achieve different core masses.

For the nuclear network we use the 22 isotopes of `approx21\_cr60\_plus\_co56' (e.g., \citealt{Timmes1999}), aimed at stellar evolution up to collapse. This network includes hydrogen, $\rm He3$ and $\rm He4$ up to the heavier isotopes of $\rm Fe52~,~Fe54~,~Fe56~,~Co56~,~Ni56~,~Cr60$.
%neut,h1,prot,he3,he4,c12,n14,o16,ne20,mg24,si28,s32,ar36,ca40,ti44,cr48,cr60,fe52,fe54,fe56,co56,ni56

We scale mesh refinement gradually up to `max\_dq' values of $1d-4$ at the later stages (from the default value of $1d-2$) to properly resolve the fine burning features close to core collapse.

The \mesa~equation of state (EOS) is a blend of the OPAL \citep{Rogers2002}, SCVH \citep{Saumon1995}, FreeEOS \citep{Irwin2004}, HELM \citep{Timmes2000}, PC \citep{Potekhin2010}, and Skye \citep{Jermyn2021} EOSs.
Nuclear reaction rates are from JINA REACLIB \citep{Cyburt2010}, NACRE \citep{Angulo1999} and additional tabulated weak reaction rates \citet{Fuller1985, Oda1994, Langanke2000}.  Screening is included via the prescription of \citet{Chugunov2007}.
Thermal neutrino loss rates are from \citet{Itoh1996}.
Radiative opacities are primarily from OPAL \citep{Iglesias1993, Iglesias1996}, with low-temperature data from \citet{Ferguson2005} and the high-temperature, Compton-scattering dominated regime by \citet{Poutanen2017}.  Electron conduction opacities are from \citet{Cassisi2007} and \citet{Blouin2020}.

% ==========================================================
\section{Binding energy estimation}
\label{App2:BEcalc}
% ==========================================================
\renewcommand\thefigure{B.\arabic{figure}}
\renewcommand\thetable{B.\arabic{table}}

Because of numerical difficulties of stripped-envelope progenitors (specifically some steep gradients) some simulations did not reach the phase of core collapse, although they did reach oxygen depletion and/or silicon depletion at the center. Time steps became much too short and we had to terminate the simulations before core collapse. In these cases we estimated the binding energy at collapse (red-stars in Fig. \ref{fig:BindingEnergy}) by extrapolating the binding energy during earlier phases using linear fits. 

\begin{figure*}
\begin{center}
 \includegraphics[trim=1.5cm 0cm 0cm .57cm,scale=0.65]{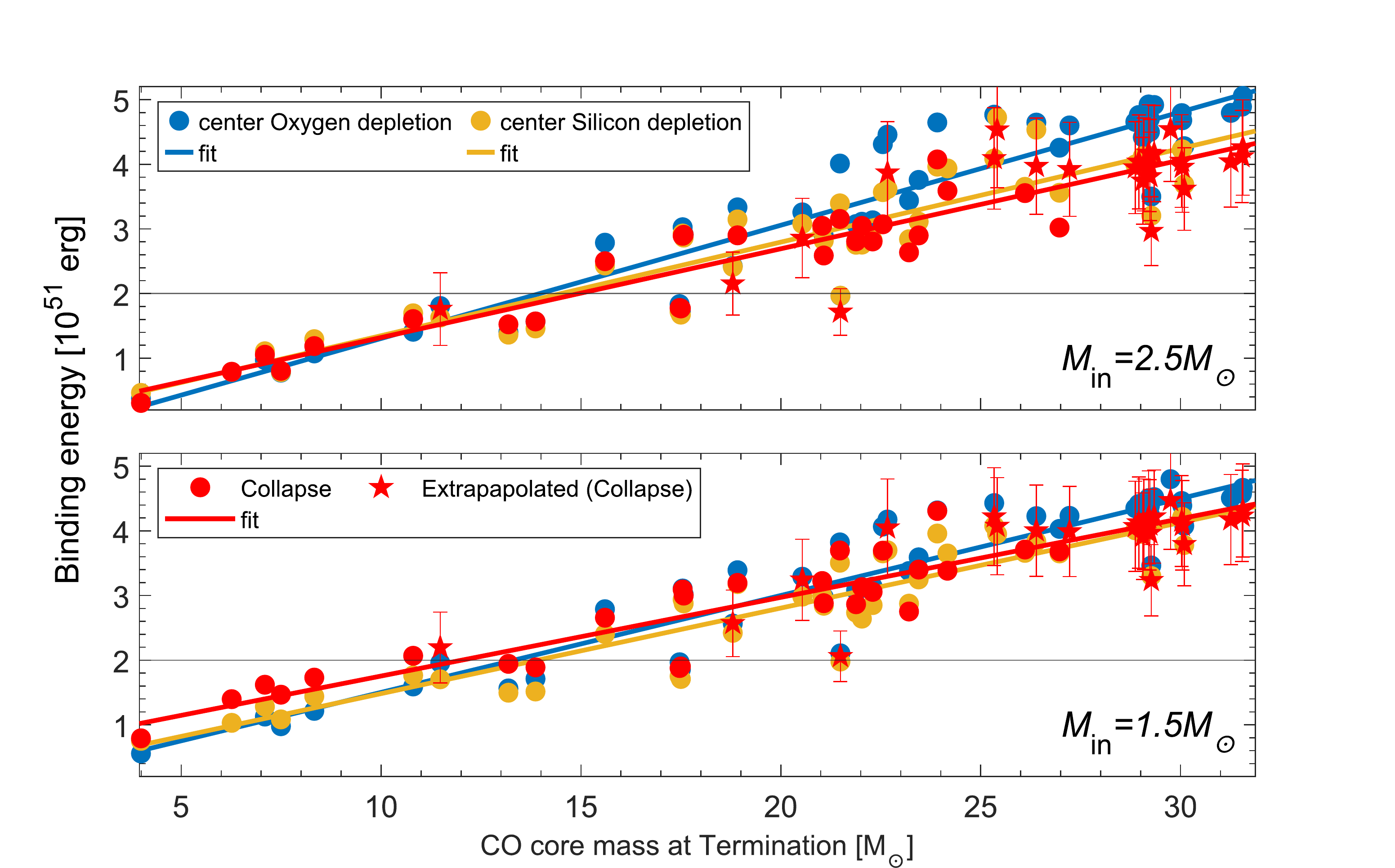}
 %1.2cm 7.2cm 1.8cm 7.85cm
\caption{The binding energy of the envelope above mass coordinate $M_{\rm in}=2.5 M_\odot$ (upper panel) and $M_{\rm in}=1.5 M_\odot$ (lower panel) as a function of the final carbon-oxygen core mass.
The blue circles are at central oxygen depletion ($5 \%$ oxygen in the center), the orange circles are at silicon depletion ($5 \%$ silicon in the center), and red circles are at core collapse. The three respective lines are the linear fit to the points. Red stars are the extrapolated values for the binding energy at collapse based on available earlier data points (oxygen depletion or silicon depletion) for the cases that did not reach collapse (see text).}
\label{fig:App3fig}
\end{center}
\end{figure*}

We made linear fits to the binding energies as function of the CO core masses at three evolutionary phases: oxygen depletion, silicon depletion, and core collapse. In Fig.~\ref{fig:App3fig} we present these three fittings by blue, orange, and red lines, respectively, for $M_{\rm in}=2.5M_\odot$ (upper panel) and $M_{\rm in} = 1.5M_\odot$ (lower panel). 
From these three lines we can find the ratio of the binding energy at core collapse to the binding energy at oxygen depletion and to the binding energy at silicon depletion. In cases where we did not reach core collapse we use this ratio at the given CO core mass to calculate the expected binding energy at core collapse. We mark these energies by red-stars in Fig. \ref{fig:App3fig} and use them in Fig. \ref{fig:BindingEnergy}. Error bars attached to the red stars signify the $1\sigma$ intervals of the this extrapolation procedure.
We note that the CO core mass does not change much after oxygen depletion in the non-extended helium phase simulations. The average difference between the CO core mass at central oxygen depletion and at core collapse is $\Delta m_{\rm CO}^{\rm core}=0.06\pm 0.33 M_\odot$.
% FFFFFFFFFFFFFFFFFFFFFFFFFFFFFFFFFFFFFFFFFFFFFFFFFFFFFFFFFFFFF
%  trim={<left> <lower> <right> <upper>}
% FFFFFFFFFFFFFFFFFFFFFFFFFFFFFFFFFFFFFFFFFFFFFFFFFFFFFFFFFFFFF

We fit the binding energy $E_{\rm bind}$ versus the CO core mass $M_{\rm CO}$ by a linear fit $E_{\rm bind} = a M_{\rm CO} + b $. In Table \ref{tab:BElines} we list the values of the two coefficients for the six lines (three stage for two values of the mass that collapses to form the NS). We also list (last column) the number of data points that were used at each fitting.
\begin{table}[!ht]
\begin{center}
\hspace*{-2.5cm}\begin{tabular}{|| c | c || c | c | c | c | c ||} 
\hline
& $a M_{\rm CO}+b$ & $\rm Fit_{1.5M_\odot}$ & $\rm Fit_{2.5M_\odot}$ & $\rm Fit_{BE break}$ & $\rm Fit_{Fe core}$ & No. of points \\ [0.5ex]  
%% PASTE FROM HERE
 \hline\hline
\multirow{2}{7em}{Collapse} 
& a $[\rm{10^{51} erg}/M_\odot]$ & $0.122 \pm 0.024$ & $0.137 \pm 0.024$ & $0.131 \pm 0.022$ & $0.124 \pm 0.023$ 
& \multirow{2}{1em}{27} \\ 
& b $[\rm{10^{51} erg}]$         & $0.537 \pm 0.46$ & $-0.053 \pm 0.444$ & $-0.028 \pm 0.421$ & $0.314 \pm 0.424$  \\
 \hline
\multirow{2}{7em}{$\rm Si_{cntr}~ depletion$} 
& a $[\rm{10^{51} erg}/M_\odot]$ & $0.132 \pm 0.017$ & $0.145 \pm 0.02$ & $ -- $ & $ -- $ 
& \multirow{2}{1em}{43} \\
& b $[\rm{10^{51} erg}]$         & $0.159 \pm 0.38$ & $-0.099 \pm 0.445$ & $ -- $ & $ -- $  \\
 \hline
 \multirow{2}{7em}{$\rm O_{cntr}~ depletion$} 
& a $[\rm{10^{51} erg}/M_\odot]$ & $0.15 \pm 0.017$ & $0.175 \pm 0.02$ & $ -- $ & $ -- $ 
& \multirow{2}{1em}{47} \\
& b $[\rm{10^{51} erg}]$         & $0.002 \pm 0.394$ & $-0.447 \pm 0.479$ & $ -- $ & $ -- $  \\
 \hline\hline
%% TO HERE
\end{tabular}
\caption{The linear fits to the lines in Fig. \ref{fig:App3fig} and the number of data points for each of the simulation groups: at collapse (second row), at center silicon depletion (third row) and center oxygen depletion (bottom row). The third and fourth columns are the fits to the binding energy $E_{\rm bind,1.5},E_{\rm bind,2.5}$ at mass coordinates $M_{\rm in}=1.5 M_\odot$ and $M_{\rm in}=2.5 M_\odot$, respectively. In the fifth column we present the linear fit to the variation of the binding energy at the black dots in Fig. \ref{fig:BindingEnergyDetailed} with the CO core mass. 
In the sixth column we present the linear fit to the variation of the binding energy at the edge of the iron core (gray squares in Fig. \ref{fig:BindingEnergyDetailed}) with the CO core mass. 
Linear fits are in units of energy $E_{\rm bind}~[\rm 10^{51} erg]$ to CO core mass $M_{\rm CO}~[\rm M_\odot]$. Values and errors ($2\sigma$) are in accordance with Fig.~\ref{fig:App3fig}.}
\label{tab:BElines}
\end{center}
\end{table}

\pagebreak

% ==========================================================
\section{Simulations parameter space}
\label{App3:simList}
% ==========================================================
\renewcommand\thefigure{C.\arabic{figure}}
\renewcommand\thetable{C.\arabic{table}}

\begin{figure*}[h!]
\begin{center}
 \includegraphics[trim=1.5cm 0cm 0cm 0cm,scale=0.65]{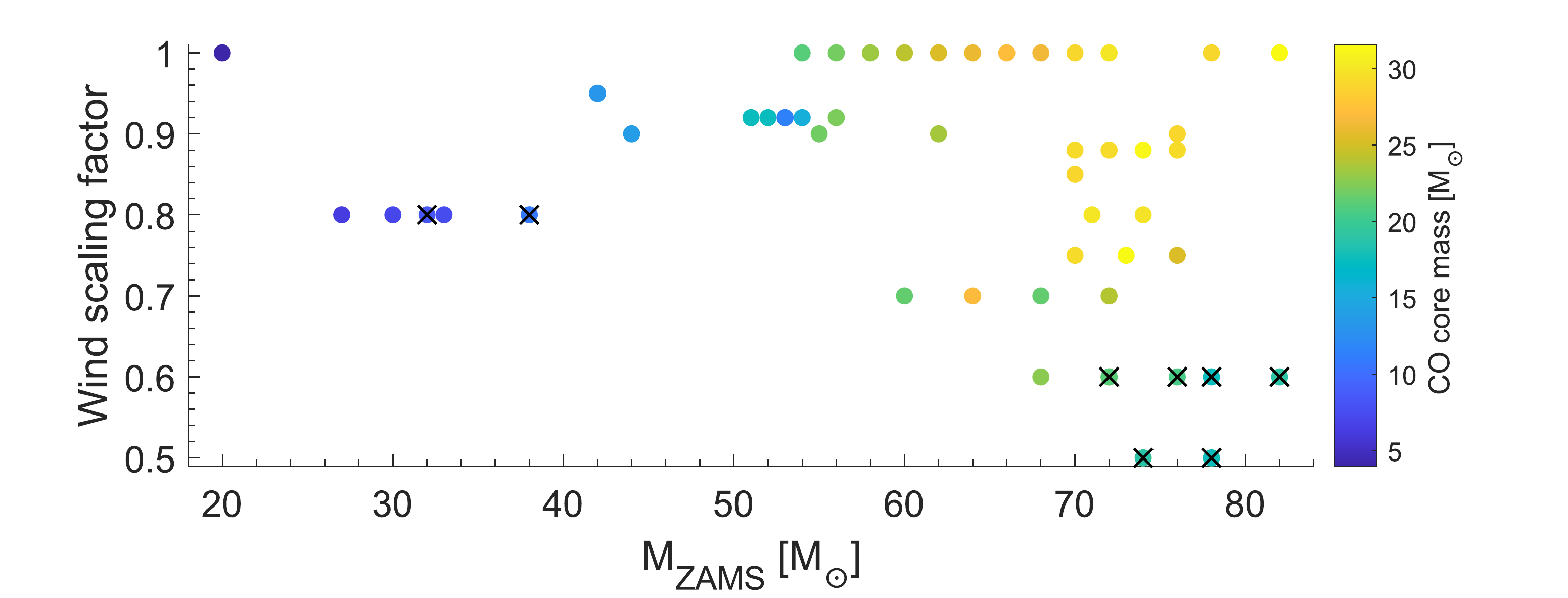}
\caption{Wind (Dutch scheme) scaling factors (vertical axis) and the ZAMS masses of the cases (horizontal axis) that we simulated, and the final CO core mass (by color bar). We denote with a black X the cases where we extended the He burning to a later stage before removing the hydrogen envelope.}
\label{fig:AppCfig}
\end{center}
\end{figure*}

In Fig.~\ref{fig:AppCfig} we present the simulations that we conducted in a three-parameters space. As input we show the dutch wind scaling factor in \mesa~in the range of $0.5 < \alpha_{\rm Dutch,wind} < 1$ and the zero age main sequence (ZAMS) mass in the range of $ 20 M_\odot < M_{\rm ZAMS} < 82 M_\odot$. As an output we present the final CO core mass (mainly oxygen mass) in units of solar mass according to the color bar. 
% FFFFFFFFFFFFFFFFFFFFFFFFFFFFFFFFFFFFFFFFFFFFFFFFFFFFFFFFFFFFF
%  trim={<left> <lower> <right> <upper>}
% FFFFFFFFFFFFFFFFFFFFFFFFFFFFFFFFFFFFFFFFFFFFFFFFFFFFFFFFFFFFF

\end{document}